\documentclass[prd,nofootinbib,floats,superscriptaddress,eqsecnum,tightenlines, showpacs,12pt]{revtex4}

\usepackage{amsfonts,amssymb,amstext,amsmath,amsthm,bm,slashed}
\usepackage[dvips]{graphicx}
\usepackage{mathtools}
\usepackage{color}
\usepackage{hyperref}

\theoremstyle{definition}

\theoremstyle{plain}

\usepackage{cleveref}
\usepackage{relsize}

\newcommand{\be}{\begin{equation}}
\newcommand{\ee}{\end{equation}}
\newcommand{\barray}{\begin{array}}
\newcommand{\earray}{\end{array}}
\newcommand{\bea}{\begin{eqnarray}}
\newcommand{\eea}{\end{eqnarray}}
\newcommand{\bs}{\begin{subequations}}
\newcommand{\es}{\end{subequations}}
\newcommand{\beal}{\begin{align}}
\newcommand{\eeal}{\end{align}}

\def\eps{\epsilon}

\def\lam{\lambda}

\def\a{\alpha}
\def\b{\beta}
\def\eps{\epsilon}

\def\bs{\bar{\sigma}}

\newcommand{\rd}{\mathrm{d}}

\def\sl2c{SL(2,\mathbb{C})}

\newcommand{\mc}[1]{\mathcal{#1}}

\newcommand{\abs}[1]{\vert #1 \vert}
\newcommand{\com}[1]{\left[#1\right]}
\newcommand{\pb}[1]{\left\lbrace #1 \right\rbrace}
\newcommand*\overbar[1]{%
  \hbox{%
    \vbox{%
      \hrule height 0.5pt 
      \kern0.4ex
      \hbox{%
        \kern 0em
        \ensuremath{#1}%
        \kern 0em
      }%
    }%
  }%
}

\newcommand{\mbf}[1]{\mathbf{#1}}
\newcommand{\mbg}[1]{\boldsymbol{#1}}
\def\bs{\bar{s}}

\newcommand{\R}{\mathbb{R}}

\usepackage{bbm}

\def\R{{\mathbbm R}}

\newcommand{\mr}[1]{\mathrm{#1}}

\allowdisplaybreaks

\usepackage{tikz}
\usetikzlibrary{decorations.pathmorphing}
\usepackage{tkz-euclide}
\usetikzlibrary{decorations.markings}
\tikzset{->-/.style={decoration={
  markings,
  mark=at position .5 with {\arrow{>}}},postaction={decorate}}}
\pgfdeclarelayer{bg}    
\pgfsetlayers{bg,main}

\begin{document}
\title{A Bilocal Model for the Relativistic Spinning Particle}

\author{{Trevor Rempel}}

\affiliation{{Perimeter Institute for Theoretical Physics, Waterloo, Ontario, Canada}}
\affiliation{\small\textit{Department of Physics, University of Waterloo, Waterloo, Ontario, Canada}}
\smallskip
\author{{Laurent Freidel}}
\smallskip 
\affiliation{{Perimeter Institute for Theoretical Physics, Waterloo, Ontario, Canada}}
\date{ \today}
\bigskip
\begin{abstract}
In this work we show that a relativistic spinning particle can be described 
at the classical and the quantum level as
being composed of two physical constituents which are entangled and separated by a fixed distance.
This bilocal model for spinning particles allows for a natural description of particle interactions as a local interaction at each of the constituents. This form of the interaction vertex provides a resolution to a long standing issue on the nature of relativistic interactions for spinning objects in the context of the worldline formalism. It also potentially brings a dynamical explanation for why massive fundamental objects are naturally of lowest spin.
We analyze first a non-relativistic system where spin is modeled as an entangled state of two particles with the entanglement encoded into a set of constraints. It is shown that these constraints can be made relativistic and that the resulting description is isomorphic to the usual description of the phase space of massive relativistic particles with the restriction that the quantum spin has to be an integer.
\end{abstract}

\maketitle

\section{Introduction}
That elementary particles might possess a finite extension has a long history, dating back to Lorentz's theory of the electron. The advent of local quantum field theory superseded these early notions, modeling elementary particles as field quanta with no internal geometry. In the 1950s, persistent divergences in the description of hadrons prompted Yukawa \cite{yukawa_1949A,yukawa_1950B} to reconsider these canonical ideas, showing that particles with an intrinsic extension could be modeled by means of a simple bilocal field theory. Unfortunately, these models possessed a number of undesirable features and ultimately fell out of favor when QCD realized an accurate description of hadrons as point like field quanta. Bilocal models would have been relegated to the history books were it not for the advent of another model which also emerged around this time. String theory began as an attempt to understand certain QCD processes and is by far the most studied model in which elementary particles are considered to have a finite extension. There is an intimate link between string theory and bilocal models, with several varieties of the latter being published \cite{takabayasi_1979,two_interacting_relativistic,second_quantization_nonhadrons} following the work of Yukawa. In particular, many of the aforementioned models can be viewed as restrictions on the motion of a classical string \cite{bilocal_and_string}. More recently bilocal models have emerged in the context of higher spin theory in an attempt to  derive the form  of interaction vertices \cite{mechanical_higher_spin}. Presently we will investigate further applications of bilocal models, showing that they play a fundamental role in our understanding of spin.\\
\indent In a recent work \cite{rempel_2015} we presented a classical model of the relativistic spinning particle which was based on an application of the coadjoint orbit method \cite{kirillov_2002} to the Poincar\'e group. This ``Dual Phase Space'' model (DPS) considered the naive phase space of a spinning particle to be parameterized by two pairs of canonically conjugate four-vectors $(x^\mu, p^\mu)$ and $(\chi^\mu, \pi^\mu)$. The former corresponded to the standard position and momentum variables while the latter encoded the spinning degrees of freedom and were subject to a constraint 
$(\chi^2/\ell^2 + \pi^2/\epsilon^2) = 2s^2$, where we needed to introduce a fundamental length $\ell$ and energy $\epsilon$ such that $\hbar =\ell\epsilon$.
These were  supplemented by orthogonality conditions $p\cdot { \pi}= { \pi}\cdot {\chi} =0$.  This structure is strikingly similar to the phase space of a two particle system subject to relativistic constraints. In what follows we will formalize this observation and show that the relativistic spinning particle can be realized as a bilocal model. In other words the spinning particle can be described as being composed of two constituents {\it entangled} together by a relativistically invariant constraint. 
The constituents in questions are spinless relativistic particles which can be taken as either massive or massless.
As we will see the nature of how these physical  constituents are tied up together is not through a confining potential, but through a relativistic constraint.
It is well know that constraints generate entanglement at the quantum level, for example $J=0$ creates the EPR entanglement of two spins. 
Our conclusion is that the fundamental entanglement of two relativistic particles defines spin. As we will see, even if the constituents can be massless the resulting entangled spinning particle is massive with a mass that satisfies a bound $m^2 \geq s^2 \epsilon^2$.
The main confirmation of our results  comes from studying the relativistic interactions of spinning particles. It turns out that 
a consistent interaction between relativistic spinning particles amounts to simply demanding locality for each  constituent in the bilocal particle.\\
\indent We emphasize that a bilocal interpretation not only realizes the dual phase space model (DPS) exactly but also captures the intuition we have regarding the nature of spin. The non-relativistic model we begin with is purposefully naive, viewing a spinning particle as two point like objects, coupled by a rigid rod with a fixed angular momentum about the center of mass. As a constrained system the model is easily quantized and we show that it yields the correct values for the spin operators $\hat{S}^2$ and $\hat{S}_3$ provided that we restrict to integer spin. The desired bilocal model is then the relativistic extension of this simple non-relativistic system. This gives a completely down to earth and elementary description of relativistic spin as a relativistic rigid rod. We establish the equivalence of this description with DPS and show that the three point interaction vertex considered in  \cite{rempel_2015} is interpreted in the bilocal picture as discussed above, i.e. local interactions at the constituent particles.  \\
\indent As far as we can tell the bilocal perspective on massive spinning particles along with the detailed study of the corresponding model done here is new. However, the model is not unrelated to other bilocal models that have been explored in the literature \cite{takabayasi_1979,two_interacting_relativistic,second_quantization_nonhadrons}. In a sense our model is the  most highly constrained a two particle model can  be and we show that many of the aforementioned bilocal models can be obtained from the present one by dropping and/or combining constraints. To conclude the paper we present the quantum version of the relativistic two particle model, showing that the spin part of the wavefunction is identical to the one derived in the non-relativistic case.  

\section{Non-relativistic Two Particle Model}
\subsection{Hamiltonian Formulation}
Let's consider a system comprised of two non-relativistic point particles with masses $m_1$ and $m_2$. The corresponding phase space is parametrized by the position and momenta of each particle $(\vec{x}_1, \vec{p}_1)$ and $(\vec{x}_2,\vec{p}_2)$ with standard Poisson bracket structure
\begin{align}\label{BILOC:nonrel-poisson}
\pb{x_i^a,p_j^b} = \delta_{ij}\delta^{ab}, \qquad i,j = 1,2 \;\; \mr{and}\;\; a,b = 1,2,3.
\end{align}
Let $M = m_1 + m_2$ be the total mass of the system and $\mu = m_1m_2/M$ the reduced mass, then we can introduce:
\begin{align}\label{BILOC:com}
\vec{X} = \frac{m_1}{M}\vec{x}_1 + \frac{m_2}{M}\vec{x}_2, \qquad \Delta \vec{x} = \vec{x}_1 - \vec{x}_2,
\end{align}
where $\vec{X}$ are the coordinates of the center of mass and $\Delta \vec{x}$ is the relative displacement between the particles. Momenta conjugate to these coordinates are given by 
\begin{align}
\vec{P} = \vec{p}_1 + \vec{p}_2, \qquad \Delta \vec{p} = \frac{\mu}{m_1}\vec{p}_1 - \frac{\mu}{m_2}\vec{p}_2,
\end{align} 
respectively.
These definitions imply the following non-vanishing Poisson brackets
\begin{align}
\pb{X^a, P^b} = \delta^{ab}, \qquad \pb{\Delta x^a, \Delta p^b} = \delta^{ab}.
\end{align}
The coordinates introduced above can also be used to decompose the total angular momentum of the two particle system as the sum of the total and relative angular momenta 
\begin{align}
\vec{J} &\coloneqq \vec{x}_1\times \vec{p}_1 + \vec{x}_2 + \vec{p}_2 \\
&= \vec{X}\times \vec{P} + \Delta \vec{x} \times \Delta \vec{p}.
\end{align}
Note that the second equality shows that  $\vec{J} = \vec{L} + \vec{S}$, where $\vec{L} = \vec{X}\times \vec{P}$ is the ``external'' angular momentum associated with motion of the system as a whole while $\vec{S} = \Delta \vec{x} \times \Delta \vec{p}$ is the ``internal'' angular momentum resulting from the rotation around the center of mass. This internal rotation represents  the spin degrees of freedom.\\
\begin{figure}[h]
\begin{center}
\begin{tikzpicture}
\node[fill=black, circle, scale = 2pt, label = {south, black: $m_1$}] (p1) at (0,0) {};
\node[fill=black, circle, scale = 2pt, label = {north, black: $m_2$}] (p2) at (6,0) {};
\node[fill =none](dummy1) at (3,3){};
\node[fill= none](dummy2) at (3,-3){};
\draw [very thick](p1.east)  -- node[label = {[label distance = 0.5]north :$\ell$}]{} (p2.west);
\draw [thick, ->] (p1) edge[bend left = 45] node[label = {[label distance = 0.5]north west :$\hbar s$}]{} (dummy1);
\draw [thick, ->] (p2) edge[bend left = 45] (dummy2);
\end{tikzpicture}
\end{center}
\caption{Two particles connected by rigid rod of length $\ell$ and pictured in the center of mass frame where the total angular momentum has magnitude $\hbar s$.}
\label{BILOC:fig:setup}
\end{figure}
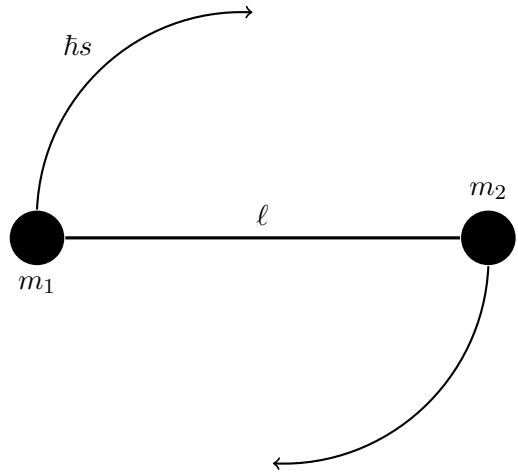
\indent At this point we have a pair of free non-relativistic particles and it remains to impose some structure on the system which will make contact with intuitions we have regarding the nature of spinning particles. Classically, a spinning particle is a rigid object with a fixed, non-zero value for its ``internal'' angular momentum. The former condition can be implemented by demanding that the two particles are coupled by a rigid rod of length $\ell$ and the latter by setting the magnitude of the angular momentum in the center of mass frame to be $\hbar s$, for some dimensionless constant $s$.  This amounts to imposing the constraints
\begin{align}\label{BILOC:constraints}
(\Delta \vec{x})^2 = \ell^2, \qquad \mr{and} \qquad (\Delta \vec{x} \times \Delta \vec{p})^2 = \hbar^2 s^2,
\end{align}    
see Figure \ref{BILOC:fig:setup}. These constraints satisfy a closed algebra. A Hamiltonian can now be constructed by adding the constraints in \cref{BILOC:constraints} to the standard Hamiltonian for a system of two free particles\footnote{A similar model appeared in a different context in \cite{non_rel_model}.}
\begin{align}\label{BILOC:non-rel Ham}
H &=  \frac{1}{2m_1}\vec{p}_1^{\,2} + \frac{1}{2m_2}\vec{p}_2^{\,2} + \frac{\lam_1}{2}\left[(\Delta \vec{x})^2 - \ell^2\right] + \frac{\lam_2}{2} \left[(\Delta \vec{x} \times \Delta \vec{p} \,)^2 - \hbar^2 s^2 \right],
\end{align}
where $\lam_1$ and $\lam_2$ are Lagrange multipliers. To ensure that the constraints are stationary under the evolution defined by $H$ we need to include $\Delta \vec{x} \cdot \Delta \vec{p} = 0$ which allows us to re-write the {\it full} Hamiltonian as
\begin{align}\label{BILOC:non-rel Ham-full}
H &= \frac{1}{2M}\vec{P}^{\,2} + \frac{1}{2\mu} (\Delta \vec{p}\,)^2 + \frac{\lam_1}{2}\left[(\Delta \vec{x})^2 - \ell^2\right] + \frac{\lam_2}{2} \left[(\Delta \vec{p} \,)^2 - \eps^2{s}^2 \right] + \lam_3\Delta \vec{x} \cdot \Delta \vec{p},
\end{align}
where $\eps$ has units of energy and satisfies $\eps \ell = \hbar$. No further constraints are required but due to the second class nature of the constraints imposed, the condition that all the constraints are preserved under time evolution imposes the following relations between Lagrange multipliers:
\begin{align}
\lam_2 = \frac{\ell^2}{\eps^2 s^2}\lam_1 - \frac{1}{\mu} \qquad \mr{and} \qquad \lam_3 = 0.
\end{align}
The final form of the non-relativistic {\it restricted} Hamiltonian is therefore, up to a constant term $ \eps^2s^2/2\mu$, given by
\begin{align}\label{BILOC:non-rel Ham-final}
H = \frac{1}{2M}\vec{P}^2 
+{\lambda}\left[\frac12\left(\frac{\Delta \vec{p}}{\eps}\right)^2 \, +  \frac{s^2}{2} \left(\frac{\Delta \vec{x}}{\ell}\right)^2  - s^2\right],
\end{align}
where $\lam = \lam_1\ell^2/ s^2$. As one can see from $H$ there is a single first class constraint
\begin{align}\label{BILOC:sho-ham}
\ell^2(\Delta \vec{p}\,)^2 + \eps^2 s^2 (\Delta \vec{x})^2  = 2\hbar^2 s^2,
\end{align}
and two second class constraints
\begin{align}
(\Delta \vec{x})\cdot (\Delta \vec{p}) = 0 \qquad \mr{and} \qquad \eps^2 s^2(\Delta \vec{x})^2 - \ell^2(\Delta \vec{p}\,)^2 = 0.
\end{align} 
The dimension of the reduced phase space is therefore $12 - 1\times 2 - 2\times 1 = 8$ for a total of $4$ physical degrees of freedom; as expected for a spinning particle (3 for position and 1 for the spin). The motion of the composite system can be deduced by examining the Hamiltonian \cref{BILOC:non-rel Ham-final}. The unconstrained part of $H$ indicates that the center of mass evolves like a free particle, while the single first class constraint is a harmonic oscillator potential acting on the relative separation, and so the latter will execute periodic motion with frequency $\omega \propto s$.

\subsection{Lagrangian Formulation}
It is a straightforward exercise to compute the Lagrangian for this model, beginning with $H$ as given in \cref{BILOC:non-rel Ham-full} we put $L = \vec{P}\cdot \dot{\vec{X}} + \Delta \vec{p} \cdot \Delta \dot{\vec{x}} - H$. 
We can now integrate out the momenta, after which the Lagrange multiplier $\lambda_3$ enters quadratically and therefore can also be integrated without difficulty.
One obtains
\be
L= \frac{M}{2}\dot{\vec{X}}^2 + \frac12 \frac{\mu}{(1+\lambda_2\mu)} (D_t \Delta \vec{x})^2 + \frac{\lambda_2}2 \epsilon^2 s^2 - \frac{\lam_1}{2}\left[(\Delta \vec{x})^2 - \ell^2\right],
\ee
where 
\begin{align}
D_t \Delta \vec{x} \coloneqq \Delta \dot{\vec{x}} - \frac{(\Delta \dot{\vec{x}} \cdot \Delta \vec{x})}{(\Delta \vec{x})^2}\Delta \vec{x},
\end{align}
is a  covariant time derivative which preserves the constraint $(\Delta \vec{x})^2 = \ell^2$. It projects the relative motion $\Delta \dot{\vec{x}}$ orthogonal to $\Delta \vec{x}$. 
The Lagrange multiplier $\lambda_2$ doesn't enter quadratically but we can still solve for it at the classical level. The solution space possesses two branches which are labelled by a sign  $\alpha:= {\mathrm{sign}(1+\lambda_2\mu)}$. Encoding this sign into the spin by $s:= \alpha |s|$, we see that the Lagrangian can be expressed purely in terms of the configuration variables and is given by $L= L_s + \frac{\lam_1}{2}\left[(\Delta \vec{x})^2 - \ell^2\right]-\frac12 \frac{\epsilon s}{\mu}$ where the spin Lagrangian is
simply 
\begin{align}\label{BILOC:reducedL}
L_s = \frac{M}{2}\dot{\vec{X}}^2 +  \eps s \abs{D_t \Delta \vec{x}}.
\end{align}
We see that the inclusion of spin amounts to a modification of the kinetic energy which is linear in the velocity instead of quadratic. The spin $s$ itself entering as a ``stiffness'' parameter multiplying the spin kinetic energy $\abs{D_t \Delta \vec{x}}$.
 The final Lagrange multiplier $\lam_1$ imposes the constraint $(\Delta \vec{x})^2 = \ell^2$ which can be solved by introducing new variables $\vec{y}$ defined implicitly via
\begin{align}
\Delta \vec{x} = \frac{\ell}{\abs{\vec{y}}}\vec{y}.
\end{align} 
The Lagrangian \cref{BILOC:reducedL} then becomes
\begin{align}\label{BILOC:yLagrangian}
L = \frac{M}{2}\dot{\vec{X}}^2 + \frac{ \hbar s}{\abs{\vec{y}\,}} \abs{\mc{D}_t \vec{y}\,}-\frac12 \frac{\epsilon s}{\mu},
\end{align}
where $\mc{D}_t\vec{y}$ is the derivative $\dot{\vec{y}}$ projected orthogonally to $\vec{y}$. It satisfies $\mc{D}_t(\rho\vec{y})=\rho \mc{D}_t\vec{y}$. Notice that the reduced mass enters only in an overall constant factor.

\subsection{Quantizing the Non-Relativistic Model}
In this section we will quantize the non-relativistic model and show that it reproduces the expected results for a non-relativistic spinning particle. Start with the Lagrangian \cref{BILOC:yLagrangian} and compute the momenta conjugate to $\vec{X}$ and $\vec{y}$, viz
\begin{align}
\vec{P}_X = M\dot{\vec{X}},\qquad \vec{P}_y = \frac{\hbar s}{\abs{\vec{y}\,}\abs{\mc{D}_t\vec{y}\,}} \mc{D}_t\vec{y}.
\end{align}
It is straightforward to verify that $\vec{P}_y$ satisfies the constraints
\begin{align}\label{BILOC:ycon}
\vec{P}_y \cdot \vec{y} = 0, \qquad \vec{P}_y^2- \frac{\hbar^2 s^2}{\vec{y}^2}= 0,
\end{align}
and so the Hamiltonian is given as
\begin{align}
H = \frac{\vec{P}_X^2}{2M} + \lam_1\left(\vec{P}_y \cdot \vec{y}\right) + \frac{\lam_2}{2}\left(\vec{P}_y^2- \frac{\hbar^2 s^2}{\vec{y}^2}\right).
\end{align}
The Poisson brackets are standard
\begin{align}\label{BILOC:ybrackets}
\pb{X_i, P_X^j} = \delta_i^j \qquad \pb{y_i, P_y^j} = \delta_i^j
\end{align}
and can be used to show that the constraints \cref{BILOC:ycon} are first class.\\
\indent The absence of second class constraints in conjunction with \cref{BILOC:ybrackets} implies that we can quantize by making the standard replacements
\begin{gather}
\hat{X}_i\Psi = X_i\Psi,\qquad \hat{P}_X^i\Psi = -i\hbar \frac{\partial}{\partial X_i}\Psi,\\
\hat{y}_i\Psi = y_i\Psi, \qquad \hat{P}_y^i\Psi = -i\hbar \frac{\partial}{\partial y_i}\Psi,
\end{gather}
where $\Psi = \Psi(\vec{X}, \vec{y},t)$. Observe that the unconstrained part of $H$ acts only on the variables $\vec{X}$ while the constraints act only on the $\vec{y}$. This suggests that we separate variables $\Psi(\vec{X}, \vec{y},t) = \Psi_1(\vec{X},t)\Psi_2(\vec{y})$, then the condition $H\Psi = i\hbar \partial_t\Psi$ splits into three differential equations
\begin{align}
-\frac{\hbar^2}{2M}\nabla_X^2\Psi_1&= i\hbar \frac{\partial\Psi_1}{\partial t},  \\
\sum_i y^i \frac{\partial \Psi_2}{\partial y^i} &= 0,\label{BILOC:no r}\\
\nabla^2_y\Psi_2 + \frac{s^2}{\vec{y}^2} \Psi_2&= 0.\label{BILOC:harmonics}
\end{align} 
The first equality is just Schr\"odinger's equation for a free particle indicating that the internal variables continue to evolve as a free particle even in the quantum theory. The remaining equations correspond to the first class constraints imposed on the internal variables and are most easily solved by switching to spherical coordinates. Make the replacements $\vec{y} = (r\sin\theta \cos \phi, r\sin\theta \sin\phi, r\cos\theta)$ and $\Psi_2(\vec{y}) = \psi(r, \theta , \phi)$, then equation \cref{BILOC:no r} becomes
\begin{align*}
r\frac{\partial \psi}{\partial r} = 0 \quad \Longrightarrow \qquad \psi(r, \theta, \phi) = \psi(\theta, \phi)
\end{align*}  
and so $\psi$ doesn't depend on $r$. The remaining equation \eqref{BILOC:harmonics} now takes the form
\begin{align}\label{BILOC:fermionic}
\Delta \psi:=\frac{1}{\sin\theta}\frac{\partial}{\partial \theta}\left(\sin\theta \frac{\partial \psi}{\partial \theta}\right) + \frac{1}{\sin^2\theta}\frac{\partial^2 \psi}{\partial \phi^2} = -s^2\psi.
\end{align}
Here $\Delta$ is the Laplacian on the unit sphere $S^2$ spanned by $\Delta \vec{x}/|\Delta \vec{x}|$. It is well known that the solutions of this equation for functions on the sphere are given by the so called Spherical Harmonics, which represent {\it integer } spins\footnote{
As discussed in Appendix \ref{BILOC:app:harmonics}, the most general solution of this equation which is regular for $\theta \in [0, \pi]$ and $\phi \in [0,2\pi]$ are the fermionic spherical harmonics $Y^m_\ell$ for $\ell \in \frac{\mathbb{N}}{2}$, see \cite{fermi_harmonics1,fermi_harmonics2,single_valued}. In this case however the functionals cannot be understood as depending continuously  on the sphere variables $\Delta x$.}:
\begin{align}
\psi(\theta, \phi) = Y^m_\ell(\theta, \phi), \qquad \ell \in \mathbb{N}, \quad m = -\ell, -\ell +1, \ldots, \ell -1, \ell,
\end{align}
where $s^2 = \ell(\ell +1)$. The ``internal'' angular momentum (spin) operator is $\hat{\vec{S}} = \hat{\vec{y}} \times \hat{\vec{P}}_y$ and one can verify that
\begin{align}
\hat{S}_3Y^m_\ell = m\hbar Y^m_\ell \qquad \mr{and} \qquad S^2Y^m_\ell = \hbar^2\ell(\ell + 1)Y^m_\ell,
\end{align}
which is precisely the expected result. 
Overall the total wave function is given by 
\be
\Psi(x_1,x_2) =  \Psi_1(x_1+x_2) Y\left(\frac{x_1-x_2}{|x_1-x_2|}\right)\delta(|x_1-x_2|-\ell).
\ee
This wave function cannot be split into a product $\phi_1(x_1) \phi_2(x_2)$ showing that the two constituents are fundamentally entangled by the spin constraint. 
The scalar product between such functions is simply given by $|\!|\Psi|\!|^2=\int_{\mathbb{R}^3} \rd^3 x |\psi_1|^2(x) \int_{S^2}\rd n |Y|^2(n)$.

\section{Relativistic Two Particle Model}\label{BILOC:sec:rel-twoparticle}
\begin{figure}[b!]
\begin{center}
\includegraphics[scale = 0.7]{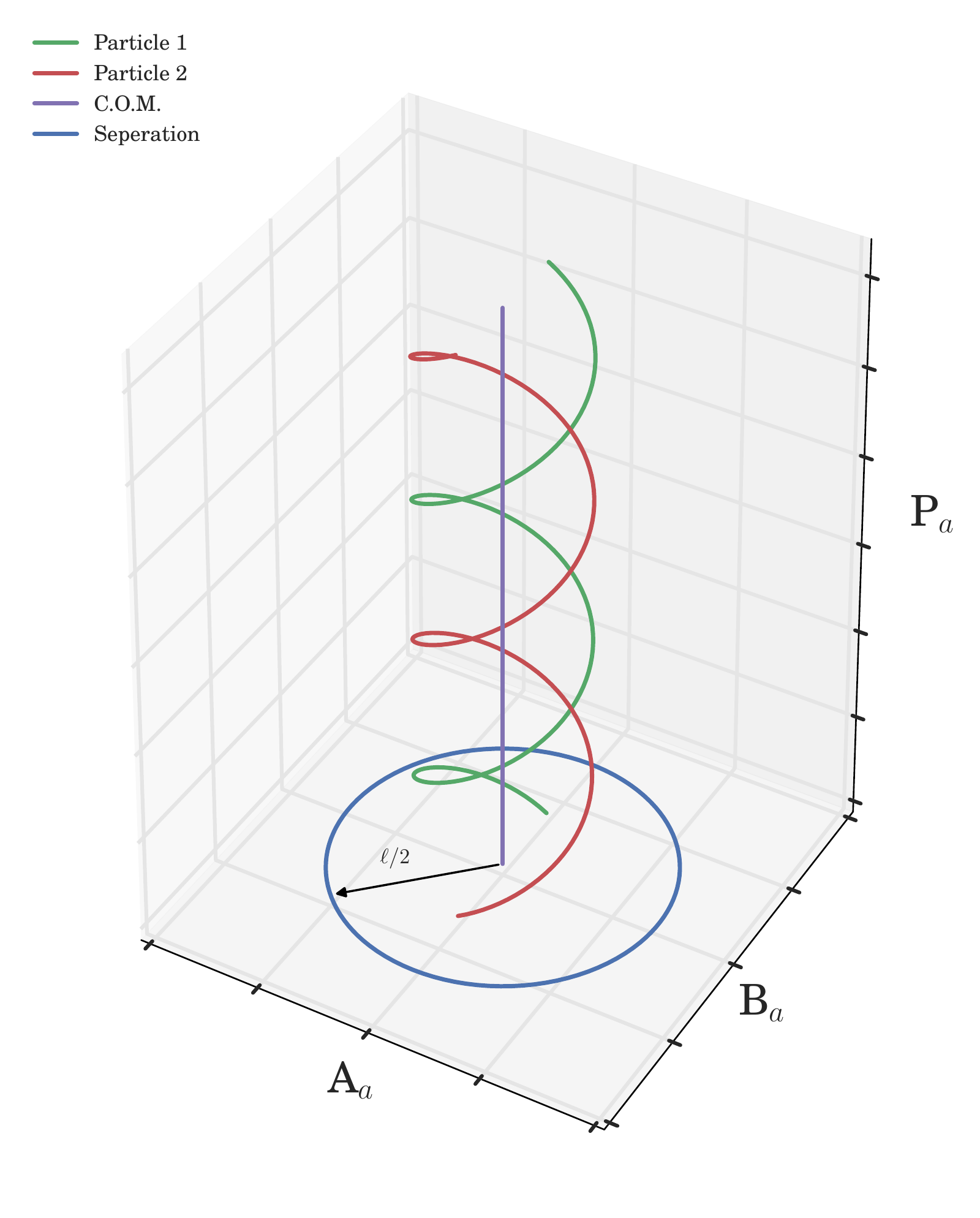}
\end{center}
\vspace{-0.75cm}
\caption{Particle trajectories plotted over two periods in the hyper-plane defined by the triplet of orthogonal vectors $(A,B,P_0)$.}
\label{BILOC:fig:motion}
\end{figure}

\begin{figure}[h]
\begin{center}
\begin{tikzpicture}
\node[fill=black, circle, scale = 0.5pt, label = {north: $x_1$}] (point1) at (1,0) {};
\node[fill=black, circle, scale = 0.5pt, label = {south: $x_2$}] (point2) at (-1,0) {};
\draw[->, thin, black] (0,-2) --(0,2)node[label={[label distance = 0.0cm]west: $B$}]{} ;
\draw[->, thin, black] (-2,0) to (2,0)node[label={[label distance = 0.0cm]south: $A$}]{};
\draw[->, thick, red] (point2) -- node[label={[label distance = 0.1cm]west: $p_2$}]{}(-1,1.5);
\draw[->, thick, red] (point1) -- node[label={[label distance = 0.1cm]east: $p_1$}]{}(1,-1.5);

\node[fill=black, circle, scale = 0.5pt, label = {south: $x_1$}] (point1) at (6,0) {};
\node[fill=black, circle, scale = 0.5pt, label = {south: $x_2$}] (point2) at (4,0) {};

\draw[->, thin, black] (5,-2) --(5,2)node[label={[label distance = 0.0cm]west: $P_0$}]{} ;
\draw[->, thin, black] (3,0) to (7,0)node[label={[label distance = 0.0cm]south: $A$}]{};

\draw[->, thick, blue] (point2) -- node[label={[label distance = 0.1cm]west: $p_2$}]{}(4,1);
\draw[->, thick, blue] (point1) -- node[label={[label distance = 0.1cm]east: $p_1$}]{}(6,1);

\node[fill=black, circle, scale = 0.5pt, label = {south east: $x_1$}] (point1) at (10,0) {};
\node[fill=black, circle, scale = 0.5pt, label = {south west: $x_2$}] (point2) at (10,0) {};
\node[label = {[label distance = 0.01cm]north: \tiny $\theta$}]  at (9.9,0.15) {};

\draw[->, thin, black] (10,-2) --(10,2)node[label={[label distance = 0.0cm]west: $P_0$}]{} ;
\draw[->, thin, black] (8,0) to (12,0)node[label={[label distance = 0.0cm]south: $B$}]{};

\draw[->, thick, green!30!black] (point1) -- node[label={[label distance = 0.1cm]west: $p_2$}]{}(9,1.5);
\draw[->, thick, green!30!black] (point1) -- node[label={[label distance = 0.1cm]east: $p_1$}]{}(11,1.5);
\draw[-,  thin, black!70!green, bend left =180] (9.8,0.3) arc (125:60:0.36cm and 0.36cm);
\end{tikzpicture}
\end{center}
\caption{Projections, at $\tau = 0$, of $(x_1,p_1)$ and $(x_2,p_2)$ into the indicated planes. The angle $\theta$ is given  in \cref{BILOC:theta}.}
\label{BILOC:fig:projection}
\end{figure}
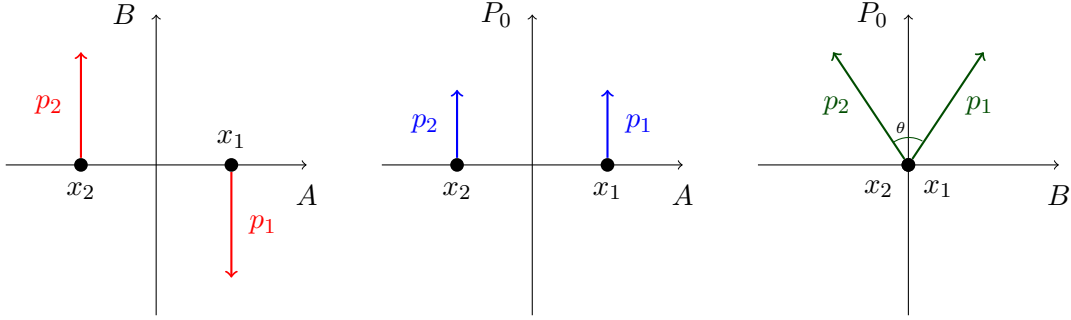

The non-relativistic model presented in the previous section captures our intuition of how a spinning particle should behave, but a  truly viable description needs to be relativistic. We begin by replacing the position and momentum variables with their four-vector counterparts $\vec{x}_i \to x^\mu_i$ and $\vec{p}_i \to p^\mu_i$, now assumed to be functions of some auxiliary parameter $\tau$. These have the standard transformation properties under elements of the Poincar\'e group $(\Lambda, y)$
\begin{align}
x_i \to \Lambda x_i + y \qquad \mr{and} \qquad p_i \to \Lambda p_i,
\end{align}
where $\Lambda$ is a Lorentz transformation and $y$ a translation. There is also a natural extension of the Poisson bracket structure in equation \cref{BILOC:nonrel-poisson} to
\begin{align}
\pb{x_i^\mu, p_i^\nu} = \delta_{ij}\eta^{\mu\nu}, \qquad i,j = 1,2, 
\end{align} 
where $\eta = \mr{diag}(-1, 1, 1, 1)$. As in the previous section we can introduce ``center of mass''\footnote{The center of mass is not a relativistically invariant quantity, hence the use of inverted commas.} and relative displacement coordinates. In doing so it will be convenient to specialize to the case where the particles are of equal mass $m_1 = m_2 = m$, whence
\begin{align}\label{BILOC:rel_var}
\begin{aligned}
X^\mu &= \frac{1}{2}\left(x_1^\mu + x_2^\mu\right), \\
P^\mu &= p_1^\mu + p_2^\mu,
\end{aligned}
\qquad\qquad
\begin{aligned}
\Delta x^\mu &= x_1^\mu - x_2^\mu,\\
\Delta p^\mu &= \frac{1}{2}\left(p_1^\mu - p_2^\mu\right).
\end{aligned}
\end{align}
Surprisingly, the case of unequal masses is significantly more complex than in the non-relativistic case and since it is not relevant for the bulk of our current analysis we have relegated its treatment to Appendix \ref{BILOC:app:unequal}. The variables in \cref{BILOC:rel_var} transform under the Poincar\'e group as
\begin{align}
X \to \Lambda X + y, \qquad P,\, \Delta p,\, \Delta x \to \Lambda P,\, \Lambda \Delta p,\, \Lambda \Delta x,
\end{align}
and one can check that $(X^\mu, P^\mu)$ and $(\Delta x^\mu, \Delta p^\mu)$ form canonically conjugate pairs. The total angular momentum $\vec{J} = \vec{L} + \vec{S}$ is generalized to an anti-symmetric tensor $J_{\mu\nu} = L_{\mu\nu} + S_{\mu\nu}$ with
\begin{align}\label{BILOC:rel-angmom}
L_{\mu\nu} = (X\wedge P)_{\mu\nu} \quad \mr{and} \quad S_{\mu\nu} =  (\Delta x \wedge \Delta p)_{\mu\nu},
\end{align}
where $(A\wedge B)_{\mu\nu} = A_\mu B_\nu - A_\nu B_\mu$. Again $L_{\mu\nu}$ represents the ``external'' angular momentum of the system as whole while $S_{\mu\nu}$ represents ``internal'' rotations. \\
\indent The relativistic Hamiltonian is a straightforward generalization of the non-relativistic one, see \cref{BILOC:non-rel Ham-final}, in particular the restricted Hamiltonian is 
\begin{align}\label{BILOC:rel-ham final}
H = \frac{N}{2\eps}\left[P^2 +4(m^2 + \eps^2s^2)\right] + {\widetilde{N}}
\left[\frac12\left(\frac{\Delta p}{\eps}\right)^2 \, +  \frac{s^2}{2} \left(\frac{\Delta x}{\ell}\right)^2  - s^2\right],
\end{align}
where $N$ and $\widetilde{N}$ are Lagrange multipliers.\\
\indent To see how \cref{BILOC:rel-ham final} comes about return to the non-relativistic Hamiltonian \cref{BILOC:non-rel Ham}. In the relativistic theory the free part becomes two mass shell constraints, recall that we are assuming particles of equal mass
\begin{align}\label{BILOC:mass-shell}
\frac{1}{2m}\vec{p}_i^{\, 2} \to (p_i^2 + m)^2, \qquad i = 1,2.
\end{align}
Each of these defines an evolution that must preserve the other two constraints \cref{BILOC:constraints}, now written as
\begin{align}\label{BILOC:primary-con}
(\Delta x)^2 = \ell^2 \qquad \mr{and} \qquad (\Delta x \wedge \Delta p)^2 =  \hbar^2s^2.
\end{align}
We can still interpret the first constraint as a rigidity condition, although now it fixes the spacetime interval between the two particles. Similarly, the second constraint can be seen as fixing the square of the ``internal'' angular momentum tensor, see equation \cref{BILOC:rel-angmom}. To ensure that both constraints are stationary, under the time evolution of each constituent, we need to include $p_1\cdot \Delta x = 0$ and $p_2 \cdot \Delta x = 0$, which then allows us to write the relativistic Hamiltonian as the following sum of six constraints
\begin{equation}\label{BILOC:rel-ham}
\begin{aligned}
H &= \frac{N_1}{2}\left(p_1^2 + m^2\right) +\frac{N_2}{2}(p_2^2 + m^2)  + \frac{\lam_1}{2}\left((\Delta x)^2 - \ell^2 \right) \\
&\qquad + \frac{\lam_2}{2}\left((\Delta p)^2 - \eps^2 s^2\right) + \lam_3(p_1\cdot \Delta x) + \lam_4(p_2\cdot \Delta x).
\end{aligned}
\end{equation}
No further constraints need to be added but demanding that the existing constraints Poisson commute with $H$ imposes the following conditions among the Lagrange multipliers
\begin{align}
\lam_3 = \lam_4 = 0,  \qquad N_1 = N_2, \qquad \lam_2 = \frac{\ell^2}{\eps^2 s^2}\lam_1 - (N_1+N_2).
\end{align}
After making these substitutions in \cref{BILOC:rel-ham} we obtain the Hamiltonian presented at the outset of this section, see \cref{BILOC:rel-ham final}. As can be easily verified, the relativistic model possesses two first class constraints
\begin{align}\label{BILOC:firstclass}
\Phi_\mc{M} = P^2 + 4(m^2 + \eps^2 s^2),\qquad \Phi_{S} = \ell^2(\Delta p)^2 + \eps^2s^2(\Delta x)^2 -2\hbar^2s^2,
\end{align}
and four second class constraints
\begin{align}
P\cdot \Delta x = 0, \quad \Delta p \cdot \Delta x = 0, \quad P\cdot \Delta p = 0, \quad \ell^2(\Delta p)^2 - \eps^2 s^2 (\Delta x)^2 = 0.
\end{align} 
Thus, the reduced phase space has dimension $16 - 2\times 2 - 4\times 1 = 8$ yielding $4$ physical degrees of freedom, as in the non-relativistic model. Note that the primary constraints \cref{BILOC:primary-con} are identical to those considered in the previous section if one transforms to the rest frame of the ``center of mass'' $P=(m,\vec{0})$ and implements $P\cdot \Delta x = P \cdot \Delta p = 0$. \\
\indent The equations of motion are obtained from Hamilton's equation $\dot{A} = \pb{H,A}$, we find 
\begin{equation}
\begin{aligned}
\frac{dX^\mu}{d\tau} &= -NP^\mu, \\
 \frac{d P^\mu}{d\tau} &= 0,
\end{aligned}
\qquad\qquad
\begin{aligned}
\frac{d \Delta x^\mu}{d\tau} &= -\widetilde{N}\ell^2 \Delta p^\mu,\\
\frac{d \Delta p^\mu}{d\tau} &= \widetilde{N}\eps^2s^2\Delta x^\mu,
\end{aligned}
\end{equation}
which are easily integrated to give
\begin{equation}\label{BILOC:EOM}
\begin{aligned}
X^\mu(\tau) &= X^\mu_0 - N\tau P_0^\mu, \\
P^\mu(\tau) &= P_0^\mu,
\end{aligned}
\qquad\qquad
\begin{aligned}
\Delta x^\mu(\tau) &= \ell\left[A^\mu\cos(\Omega \tau) + B^\mu\sin(\Omega \tau)\right],\\
\Delta p^\mu(\tau) &= \eps s \left[A^\mu\sin(\Omega \tau) - B^\mu\cos(\Omega \tau)\right],
\end{aligned}
\end{equation}
where $\Omega = \widetilde{N}\hbar s$. The constant vectors $A^\mu$, $B^\mu$ and $P^\mu_0$ satisfy $A^2 = B^2 = 1$, $P_0^2 = 4(m^2 + \eps^2 s^2)$ and $A \cdot P_0 = B\cdot P_0 = A\cdot B= 0$. As we can see, the ``center of mass'' propagates as a free particle while the relative displacement executes circular motion with frequency $\Omega$. This result conforms with our intuition about the system since in the original set-up both particles were free but constrained to rotate with constant ``internal'' angular momentum. The angle between $p_1$ and $p_2$, denoted $\theta$, can be computed from
\begin{align}\label{BILOC:theta}
p_1\cdot p_2 = -\abs{p_1}\abs{p_2}\cosh\theta \qquad \Longrightarrow \qquad \cosh \theta = 1 + \frac{2\eps^2s^2}{m^2}.
\end{align} 
The evolution is pictured in Figure \ref{BILOC:fig:motion}. In Figure \ref{BILOC:fig:projection} we plot the position and momentum of each particle at $\tau = 0$ projected into the planes defined by $\pb{A,B}$, $\pb{A,P_0}$ and $\pb{B,P_0}$. Both figures assume $X_0 = 0$. This completes our construction of a bi-local model, its relation to the relativistic spinning particle will be explored in the subsequent section.
\section{Re-interpreting the Model}
As the analysis in the previous section made apparent, the most natural variables for describing this two particle system are not the individual coordinates $(x_1,p_1)$ and $(x_2, p_2)$ but rather the ``center of mass'' $(X,P)$ and the relative displacement $(\Delta x, \Delta p)$. This suggests that we could re-interpret the model as a \textit{single} particle whose trajectory is determined by $(X,P)$ but which possesses internal degrees of freedom described by $(\Delta x, \Delta p)$. This re-interpretation is more than just a curiosity, it is an exact realization of the relativistic spinning particle. \\
\indent The ``Dual Phase Space'' Model (DPS) developed in \cite{rempel_2015}, provides a classical realization of the relativistic spinning particle by means of the coajoint orbit method \cite{kirillov_2002}. In particular, the naive phase space is parameterized by two pairs of canonically conjugate four-vectors, $(\mathbf{x}^\mu, \mathbf{p}^\mu)$ which describe the position and linear momentum of the particle and $(\mbg{\chi}^\mu, \mbg{\pi}^\mu)$ which encode the internal degrees of freedom associated with the spin. Note that we use bold faced characters to denote quantities originating in the DPS model. The Poisson brackets are trivial $\pb{\mbf{p}^\mu, \mbf{x}^\nu} = \eta^{\mu\nu}$ and $\pb{\mbg{\pi}^\mu, \mbg{\chi}^\nu} = \eta^{\mu\nu}$ while transformations under elements of the Poincar\'e group $(\Lambda, y)$ are given by
\begin{align}
\mbf{x} \to \Lambda \mbf{x} + y \qquad \mr{and} \qquad \mbf{p},\,\mbg{\pi},\,\mbg{\chi} \to \Lambda \mbf{p}, \,\Lambda \mbg{\pi}, \,\Lambda \mbg{\chi}.
\end{align}
The dynamics of DPS are defined by two first class and four second class constraints, given respectively by
\begin{gather}
\mbf{p}^2 = -\mbf{M}^2, \quad \mbg{\lam}^2\mbg{\pi}^2 + \mbg{\eps}^2\mbf{s}^2\mbg{\chi}^2 = 2\hbar^2\mbf{s}^2,\\
\mbf{p}\cdot \mbg{\pi} =0,\quad\mbf{p}\cdot \mbg{\chi} = 0,\quad \mbg{\pi} \cdot \mbg{\chi} = 0,\quad \mbg{\lam}^2\mbg{\pi}^2 - \mbg{ \eps}^2 \mbf{s}^2\mbg{\chi}^2 = 0,
\end{gather}
where $\mbf{m}$ and $\mbf{s} $ are the mass and spin of the particle while $\mbg{\eps}$ and $\mbg{\lam}$ are arbitrary energy and length scales which satisfy $\mbg{\eps\lam} = \hbar$. Comparing DPS to the relativistic two particle model presented in Section \ref{BILOC:sec:rel-twoparticle} shows an exact match under the following identifications
\begin{equation}\label{BILOC:map}
\begin{aligned}
\mbf{p} &= P, \\
\mbg{\pi}&= \Delta p,
\end{aligned}
\quad\quad
\begin{aligned}
\mbf{x} &= X,\\
\mbg{\chi} &= \Delta x,
\end{aligned}
\quad \quad
\begin{aligned}
\mbg{\eps} &= \eps, \\
\mbg{\lam} &= \ell,
\end{aligned}
\quad\quad
\begin{aligned}
\mbf{s} &= s,\\
\mbf{M}^2 &= 4(m^2 + s^2\eps^2).
\end{aligned}
\end{equation}
It is particularly interesting to note that the mass of the spinning particle 
$\mbf{m}$ is larger than the sum of the constituent masses. A mass defect is the hallmark of a confined system, but that is not what we have here. Instead there is a mass surplus, confirming the presence of entanglement with the entangled state having a higher energy than the sum of its constituents. The extra energy is exactly the energy present in the spin motion; it is given 
by $\hbar s/ \ell$ and can be lowered by having the pairs separate. 
Consequently, this constituent picture suggests that massive  particles of higher integer spin are unstable and it is energetically favored to lower the spin towards a spinless particle. A conclusion  not contradicted by nature.

We also see that the limit $m \to 0$ of massless constituents can be taken without incident, in which case the entire mass of the spinning particle arises as ``entanglement energy'' from the spin constraint. In this limit the particle radius can be expressed as
\begin{align}
r = \frac{\ell}{2} = \frac{\hbar s}{\mbf{M}},
\end{align}
which scales inversely with the mass in the same manner as the Bohr radius of an atom. The limit of massless constituent particles also provides a possible resolution to a long standing problem regarding the center of mass of a spinning particle. The center of mass of an extended rotating object is not relativistically invariant and any classical model of spin which views a spinning particle as possessing some non-zero extension encounters this problem, see \cite{costa_2012, costa_2014} for a detailed analysis. In the case of massless constituent particles this is a moot point since a system of massless particles does not have a center of mass and one is forced to consider the geometric centroid instead, which is precisely what $X^\mu$ is in the relativistic case.\\
\indent If we assume physical constituents with positive mass square, the bilocal model can only described particles whose mass is greater than its spin, since we have the relationship
\be
\mbf{M}^2 = \frac{4\hbar^2 s^2}{\ell^2} + 4m^2.
\ee
If the mass of the constituents are fixed this gives rise to a trajectory which is similar in spirit but different in details from a Regge trajectory where the  mass square is linearly related to the spin $\mbf{M}^2\geq  \alpha' J +\beta$.
To go beyond the bound $\mbf{M}\geq  \frac{2 \hbar s}{\ell}$ and describe massless particles $\mbf{M} = 0$ requires that the constituents be tachyons with $m^2 = -\hbar^2 s^2/\ell^2$.

\section{Interactions}
Given the mapping \cref{BILOC:map} between DPS and the two particle model, results from \cite{rempel_2015} can be imported directly and re-interpreted in the two particle picture. For example, interaction with a background electromagnetic field is achieved via the minimal coupling prescription
\begin{align*}
p_1 \to p_1 + \frac{q}{2}A(x_1 + x_2) \qquad p_2 \to p_2 + \frac{q}{2}A(x_1+x_2),
\end{align*} 
where $q$ is the total charge of the spinning particle. It follows that each constituent particle carries half the total charge while the electromagnetic field couples to the center of mass coordinate $X^\mu$. 
This formulation also suggest that one could investigate a generalisation of the coupling of electromagnetism to spinning particles where the location of the field interaction for the constituents 1 and 2 are not the same.\\
\indent Interactions between spinning particles were a focal point of \cite{rempel_2015} with the paper culminating in the formulation of a necessary and sufficient condition for a consistent three-point vertex. In detail, suppose a vertex has one incoming and two out going particles with coordinates $(\mbf{x}_i,\mbf{p}_i)$, $(\mbg{\pi}_i,\mbg{\chi}_i)$, where $i = 1,2,3$ and it is assumed that particle $\#1$ is incoming. The vertex is governed by conservation of linear and angular momentum along with the requirement that interactions are local in space-time, i.e. $\mbf{x}_1 = \mbf{x}_2 = \mbf{x}_3$. It turns out that consistency is possible if and only if there exists a choice of $\mbg{\chi}$ variables such that the interaction is also local in the dual space. That is one has to impose  $\mbg{\chi}_1 = \mbg{\chi}_2 = \mbg{\chi}_3$, a condition we referred to as ``dual locality''. The conservation equations then become
\begin{align}\label{BILOC:dual mom}
\mbf{p}_1 = \mbf{p}_2 + \mbf{p}_3 \qquad \mr{and} \qquad \mbg{\pi}_1 = \mbg{\pi}_2 + \mbg{\pi}_3,
\end{align}
which can be solved by elementary methods. In the two particle picture these notions have concrete interpretations: Locality plus ``dual locality'' become the condition that interactions are local for  each constituent particle, while equation \cref{BILOC:dual mom} implies conservation of momentum at each particle. This is pictured in \Cref{BILOC:fig:interaction1,BILOC:fig:interaction2,BILOC:fig:fullinter}, where we have used the notation $p_i^{(j)}$ to indicate the $i$-th constituent of particle $j$, $i = 1,2$, $j = 1,2,3$.
In \Cref{BILOC:fig:fullinter} each spinning particle is represented by a string of length $\ell$ and it is seen that the interaction splits the incoming strip into two halves.
The resulting worldsheet is not a smooth manifold but a branched 2 dimensional surface.
This form of the interaction vertex is very different from the string inspired interaction
which has been explored in the literature on massless particles \cite{mechanical_higher_spin}. 
\begin{figure}[h]
\centering
\begin{minipage}{0.45\textwidth}
\centering
\begin{tikzpicture}[scale=0.8, every node/.style={scale=0.8}]
\node[circle, fill = black!70!green,scale = 0.1pt,label = {west, black!70!green: $p^{(1)}_{2}$}] (x11) at (-0.14,-5){};
\node[circle, fill = black!70!green,scale = 0.1pt,label = {east, black!70!green: $p^{(1)}_{1}$}] (x12) at (0.14,-5){};

\node[circle, fill = red,scale = 0.1pt,label = {north east, red: $p^{(2)}_{2}$}] (x21) at (-3.2,3.4){};
\node[circle, fill = red,scale = 0.1pt,label = {south west, red: $p^{(2)}_{1}$}] (x22) at (-3.4,3.2){};

\node[circle, fill = blue,scale = 0.1pt,label = {north west, blue: $p^{(3)}_{2}$}] (x31) at (3.2,3.4){};
\node[circle, fill = blue,scale = 0.1pt,label = {south east, blue: $p^{(3)}_{1}$}] (x32) at (3.4,3.2){};

\node[fill = black, circle, scale = 2pt, label = {north:  $X$}] (xint) at (0,0){};
\draw[->, very thick,black!70!green] (x11) to (-0.14,-0.35){};
\draw[->,  very thick,black!70!green] (x12) to (0.14,-.35){};

\begin{pgfonlayer}{bg}
\draw[->, very thick, red] (0.1,0.1) to (x21);
\draw[->, very thick, red] (-0.1,-0.1) to (x22){};
\draw[->, very thick, blue] (-0.1,0.1) to (x31);
\draw[->,  very thick, blue] (0.1,-0.1) to (x32){};
\end{pgfonlayer}

\end{tikzpicture}
\caption{Three-point interaction vertex.}\label{BILOC:fig:interaction1} 
\end{minipage}\hfill
\begin{minipage}{0.45\textwidth}
\centering 
\begin{tikzpicture}[scale=0.8, every node/.style={scale=0.8}]
\node[fill = none] at (0,1.7) {};
\node[fill = none] at (0,-7.5) {};
\node[circle, scale = 0.1pt] (inter) at (0.1304,-1.1304) {};
\node[fill=black!70!green, circle, scale = 0.1pt, label = {west, black!70!green: $p^{(1)}_{2}$}] (p11) at (-2,-6) {};
\node[fill=black!70!green, circle, scale = 0.1pt, label = {east, black!70!green: $p^{(1)}_{1}$}] (p12) at (2,-6) {};
\node[fill=red, circle, scale = 0.1pt, label = {west, red:  $p^{(2)}_{2}$}] (p21) at (-4.5,-0.5) {};
\node[fill=red, circle, scale = 0.1pt, label = {west, red: $p^{(2)}_{1}$}] (p22) at (-2.5,1.5) {};
\node[fill=blue, circle, scale = 0.1pt, label = {east, blue: $p^{(3)}_{2}$}] (p31) at (2.9,1.3) {};
\node[fill=blue, circle, scale = 0.1pt, label = {north, blue:$p^{(3)}_{1}$}] (p32) at (2.9,-2.3) {};
\node[fill=black, circle, scale = 0.5pt, label = {west: $x_2$}] (int1) at (-2,-3) {};
\node[fill=black, circle, scale = 0.5pt, label = {east: $x_1$}] (int2) at (2,-3) {};
\begin{pgfonlayer}{bg}
\draw [black!70!green, decorate,decoration={brace,amplitude=10pt, mirror},xshift=-4pt,yshift=0pt] (p11) to node[label = {[label distance = 10pt]south :$\ell$}]{} (p12); 
\draw[->, very thick, black!70!green] (p11) to  (int1);
\draw[->, very thick, black!70!green] (p12) to  (int2);
\draw[->, very thick,red] (-2,-3) to (p21);
\draw[-,dashed, very thick, red] (2,-3) to (inter);
\draw[->, very thick, red] (inter) to (p22); 
\draw[-, very thick,dashed, blue] (-2,-3) to  (inter);
\draw[->, very thick, blue] (inter) to  (p31);
\draw[->, very thick, blue] (2,-3) to (p32);
\end{pgfonlayer}
\end{tikzpicture}
\vspace{-30pt}
\caption{Detailed view of three-point interaction vertex}\label{BILOC:fig:interaction2}
\end{minipage}
\end{figure}
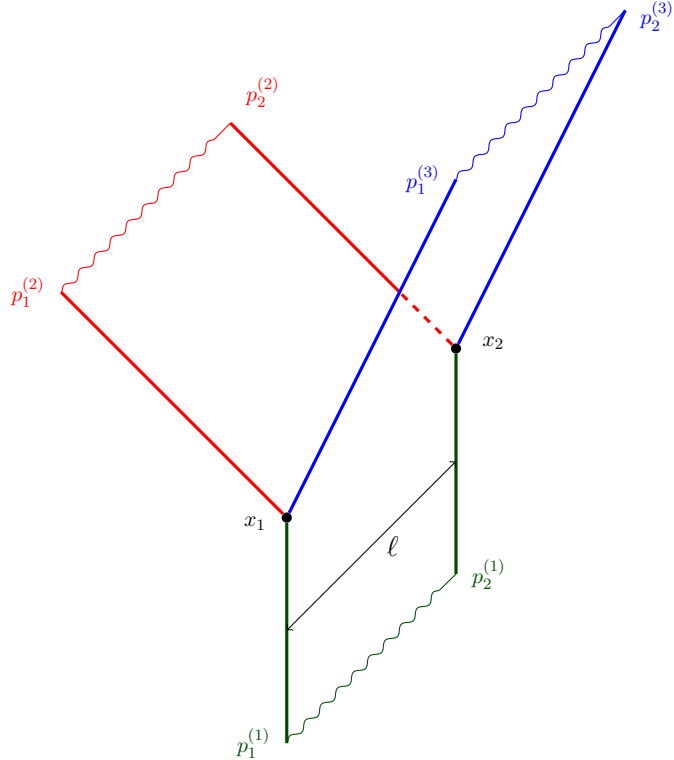
\begin{figure}
\begin{tikzpicture}[scale=0.75, every node/.style={scale=0.75}]

  \node[fill=black, circle, scale = 0.5pt] (int1) at (-2.5,-2.5) {};
  \node[label = {west: \color{black}$x_1$}] at (-2.6,-2.6) {};
  \node[fill=black, circle, scale = 0.5pt,] (int2) at (0.5,0.5) {};
  \node[label = {east: \color{black}$x_2$}] at (0.7,0.6) {};
  \node[fill = red, circle, scale = 0.1pt] (center) at (-0.5,1.5){};

  \coordinate (p11B) at ([shift = {(0,-4)}]int1)    {};
  \coordinate (p11L) at ([shift = {(0,0)}]p11B)    {};
  \draw[-,black!70!green, very thick] (p11L) -- (int1);
  \node[label = {[label distance = 0.5] west :\color{black!70!green}$p^{(1)}_1$}] at (p11L){};
  
  \coordinate (p12B) at ([shift = {(0,-4)}]int2)    {};
  \coordinate (p12L) at ([shift = {(0,0)}]p12B)    {};
  \draw[-,black!70!green, very thick] (p12L) to (int2);
  \node[label = {[label distance = 0.5] east :\color{black!70!green}$p^{(1)}_2$}] at (p12L){};

  \draw[black!70!green,decoration={snake, amplitude = .4mm,segment length = 3mm, post length=2mm},decorate]  (p11L) to (p12L);

  
  \coordinate (p21L) at ([shift = {(-4,0)}]int1)    {};
  \coordinate (p21B) at ([shift = {(0,4)}]p21L)    {};
  \draw[-,red, very thick] (int1) --  (p21B) ;
  \node[label = {[label distance = 0.5] west :\color{red}$p^{(2)}_1$}] at (p21B){};

  \coordinate (p22L) at ([shift = {(-4,0)}]int2)    {};
  \coordinate (p22B) at ([shift = {(0,4)}]p22L)    {};
  \draw[-,red, very thick, dashed] (int2) -- (center);
  \draw[red, very thick] (center) -- (p22B);
  \node[label = {[label distance = 0.5] north east :\color{red}$p^{(2)}_2$}] at (p22B){};

  \draw[red, decoration={snake, amplitude = .4mm,segment length = 3mm, post length=2mm},decorate]  (p21B) to (p22B);

  
  \coordinate (p31L) at ([shift = {(3,0)}]int1)    {};
  \coordinate (p31B) at ([shift = {(0,6)}]p31L)    {};
  \draw[-,blue, very thick] (int1)  to (p31B) ;
  \node[label = {[label distance = 0.5] west :\color{blue}$p^{(3)}_1$}] at (p31B){};

  \coordinate (p32B) at ([shift = {(3,0)}]int2)    {};
  \coordinate (p32L) at ([shift = {(0,6)}]p32B)    {};
  \draw[-,blue, very thick] (int2) to (p32L);
  \node[label={east:\color{blue}$p^{(3)}_2$}] at ([shift = {(0,-0.1)}]p32L){};
  \draw[blue, decoration={snake, amplitude = .4mm,segment length = 3mm, post length=2mm},decorate]  (p31B) to (p32L);

\draw [<->,black] (-2.5,-4.5) to node[label = {[label distance = 4pt] east :\large$\ell$}]{} (0.5,-1.5);
\end{tikzpicture}
\caption{Expanded view of three-point interaction vertex.}\label{BILOC:fig:fullinter}
\end{figure}

\section{Quantization and Other Bilocal Models}
Before examining the quantization of the relativistic two particle model it is interesting to note the relationship between DPS and other bilocal models appearing in the literature. A popular model introduced by Takabayasi \cite{takabayasi_1979} and known as the ``Simple Relativistic Oscillator Model'' (SROM) is obtained by combining  $\Phi_{\mc{M}}$ and $\Phi_S$ and dropping all remaining constraints that don't involve $P_\mu$. In particular,
\begin{align}
\Phi = \Phi_{\mc{M}} + \frac{4}{\ell^2}\Phi_S, \qquad \Phi_1 = P\cdot \Delta p, \qquad \Phi_2 = P\cdot \Delta x.
\end{align}
For a model to be interpreted as ``bilocal'' the two constituent particles need a well defined mass which means that the values of $p_i^2$ must be specified by the constraints. As $p_1,p_2 = P/2 \pm \Delta p$ we need to specify at least, $P^2 +  4(\Delta p)^2$ and $P\cdot \Delta p$. The SROM is therefore a minimally constrained bilocal model that has non-trivial kinematics in the relative separation. 

A similar model has been  proposed by Casalbuoni and Longhi \cite{second_quantization_nonhadrons}.
It imposes the 
primary constraints $P^2 + (\Delta p)^2 +(\Delta x/\alpha' )^2=0$, 
where $\alpha'$ is the inverse string tension, supplemented by $\Phi_1=\Phi_2=0$ and $(\Delta p\cdot\Delta x)=0$.
This model is obtained from a truncation of  string theory, by restricting the string motion to excite only one oscillator.
It corresponds to a limit of our model in which $m=0$, $s= 0$ and the
separation $\ell =0$ also vanish.
More precisely the relationship between the string tension and spinning particle tension is given in the limit $s\to 0$ by  $\ell^2 \sim \hbar \alpha' s^2 $.Our description does not really survive this limit since we need a non-zero separation length, so this string model is really a different model. In this limit the vertex of interaction is derived from the string vertex and has a geometry very different from the vertex we described (c.f. Figure 1 in \cite{mechanical_higher_spin}).
\\
\indent Another class of models arise by setting the total mass $\mbf{M}$ to zero or equivalently fixing $m^2 = -4\eps^2 s^2$, in which case we have tachyonic constituents. We can obtain several versions of massless higher spin particles, see the discussion by Bengtsson in \cite{mechanical_higher_spin}. 
The massless case is special, since $\mbf{M}=0$ implies that the constraints
 \be \Phi_1 = P\cdot \Delta p, \qquad \Phi_2 = P\cdot \Delta x\ee
 are first class.
 
By considering only the constraints $\Phi_\mc{M}$, $\Phi_{1}$ and $\Phi_{2}$ we obtain a theory which describes a reducible tower of higher spin massless gauge fields. Including $\Delta p \cdot \Delta x = 0$ and $\ell^2 (\Delta p)^2 = \eps^2 s^2 (\Delta x)^2 $ makes this tower irreducible and adding $\Phi_S$ as well gives a single higher spin massless gauge field. In all these models the issue of the interaction vertex is still open. 

\subsection{Quantizing the Relativistic Model}
To quantize the relativistic two particle model we will first obtain a Lagrangian description as we did in the non-relativistic case. This analysis has already been done for DPS, see eq. (50) in \cite{rempel_2015}, and since the two models are equivalent we can simply import the result. We find
\begin{align}
L_s = \eps \sqrt{\frac{s^2}{y^2}(\mc{D}_\tau y)^2 - \mbf{M}^2(\mc{D}_\tau X)^2 - \frac{2 ms}{\abs{y}}\sqrt{(\mc{D}_t X \cdot \mc{D}_t y)^2 - (\mc{D}_t X)^2(\mc{D}_t y)^2}},
\end{align}
where $\eps = \pm$ and the sign of $s$ is not fixed. These signs come from defining the square roots and
\begin{align*}
\Delta x^\mu = \ell y^\mu/\abs{y}, \quad \mc{D}_\tau A^\mu = \dot{A}^\mu - \frac{\dot{A}\cdot{y}}{y^2}y^\mu, \quad \mbf{M}^2 = 4(m^2 + \eps^2 s^2).
\end{align*}
The momenta conjugate to $X^\mu$ and $y^\mu$, denoted $P_X^\mu$ and $P_y^\mu$ respectively, can be obtained in the standard fashion by varying the action with respect to $\dot{X}$ and $\dot{y}$ respectively. There is no need to know their exact form, it is sufficient to note that they satisfy the following constraints 
\begin{gather}
P_X^2 = -\mbf{M}^2, \quad P_y^2 = \frac{s^2}{\abs{y}^2}, \quad P_y \cdot y = 0\label{BILOC:rel_first}\\
P_X\cdot y = 0, \quad P_X\cdot P_y = 0.\label{BILOC:rel_second}
\end{gather}
The first three constraints are first class\footnote{We have the standard Poisson brackets $\pb{X^\mu,P_X^\nu} =\eta^{\mu\nu}$ and $\pb{y^\mu, P_y^\nu} = \eta^{\mu\nu}$.} and are strikingly similar to those appearing in the non-relativistic model, see \cref{BILOC:ycon}. The final two constraints are second class which will complicate the quantization procedure since we must first implement Dirac brackets before promoting to commutators. Forgoing some details, we find that the commutator algebra which takes into account the second class constraints is given by
\begin{gather}
\com{\hat{X}^\mu, \hat{X}^\nu} = \frac{i}{\mbf{M}^2}\hat{S}^{\mu\nu}, \quad \com{\hat{X}^\mu, \hat{P}_X^\nu} = i\eta^{\mu\nu}, \quad \com{\hat{X}^\mu, \hat{y}^\nu} = \frac{i}{\mbf{M}^2} \hat{y}^\mu \hat{P}_X^\nu, \\
\com{\hat{X}^\mu, \hat{P}_y^\nu} = \frac{i}{\mbf{M}^2} \hat{P}_y^\mu \hat{P}_X^\nu, \quad \com{ \hat{y}^\mu,  \hat{P}_y^\nu} = i\left(\eta^{\mu\nu} + \frac{1}{\mbf{M}^2}\hat{P}_X^\mu \hat{P}_X^\nu\right),
\end{gather}
where $S_{\mu\nu}= (y\wedge p)_{\mu\nu}$ is the spin tensor and $\mbf{M}^2:= -P_X^2$.
It can be checked directly that  commutators of the second class constraints either vanish directly or are  proportional to the mass-shell constraints $(\hat{P}_X^2+ \mbf{M}^2)=0 $.\\
%
\indent Let $\mc{H} = L^2(\R^4 \times \R^4)$ be the Hilbert space of square integrable functions $\Psi(X, y)$. An action of the operators on $\mc{H}$ which respects the preceding commutation relations can be defined as follows
\begin{alignat}{3}
\hat{X}^\mu\Psi &= \left(X^\mu + \frac{i}{\mbf{M}^2}S^{\mu\nu}\frac{\partial}{\partial X^\nu}\right)\Psi, \qquad \hat{P}^\mu_X \Psi &=& -i\frac{\partial}{\partial X_\mu}\Psi,\\
\hat{y}^\mu \Psi &= \mc{P}^{\mu\nu}y_\nu\Psi, \qquad \qquad \qquad \qquad
\quad 
\hat{P}^\mu_y\Psi &=& -i\mc{P}^{\mu\nu}\frac{\partial}{\partial y^\nu}\Psi,
\end{alignat}
where
\begin{align}
S^{\mu\nu} = -i\left(y^\mu \frac{\partial}{\partial y_\nu} - y^\nu \frac{\partial}{\partial y_\mu}\right), \qquad \mc{P}^{\mu\nu} = \eta^{\mu\nu} - M^{-2}\frac{\partial^2}{\partial X_\mu \partial X_\nu}.
\end{align}
It is easily verified that the operator identities $\hat{P}_X\cdot \hat{y} = \hat{P}_X\cdot \hat{P}_y = 0$ are satisfied and so we turn our attention to the first class constraints, \cref{BILOC:rel_first}. The action of these constraints on the Hilbert space $\mc{H}$ yields the following differential equations
\begin{gather}
\Box_X\Psi = \mbf{M}^2 \Psi,\label{BILOC:klien}\\
y^\mu\frac{\partial}{\partial y_\nu}\mc{P}_{\mu\nu}\Psi = 0,\label{BILOC:rel_noR}\\
y^\mu y^\nu \frac{\partial^2}{\partial y_\a\partial y_\b}\mc{P}_{\mu\nu}\mc{P}_{\a\b}\Psi = -s^2\Psi.\label{BILOC:rel_spherical}
\end{gather}
Assuming separation of variables $\Psi(X,y) = \Psi_X(X)\Psi_y(y)$, \cref{BILOC:klien} is just the Klein-Gordon equation for $\Psi_X(X)$ which 
is easily solved in momentum space and $\Psi_X(X) =\int \rd k e^{ik\cdot X} \tilde{\Psi}_X(k) \delta(k^2+m^2)$ is the general solution.
It follows that
\begin{align}
\mc{P}^{\mu\nu}\Psi = \left(\eta^{\mu\nu} + \frac{1}{\mbf{M}^2}k^\mu k^\nu\right)\Psi \equiv \mc{P}_k^{\mu\nu}\Psi,
\end{align}
where $\mc{P}_k^{\mu\nu}$ is the projection operator onto the hyper-plane orthogonal to $k_\mu$. Let us introduce the coordinate $y_k^\mu = \mc{P}^{\mu\nu}_ky_\nu$, then we can assume a further separation of variables for $\Psi_y(y)$, namely
\begin{align}
\Psi_y(y) = \Psi_0(y\cdot k)\Psi_{y_k}(y_k).
\end{align}
We can now express \crefrange{BILOC:rel_noR}{BILOC:rel_spherical} as follows
\begin{gather}
y_k^\mu \frac{\partial}{\partial y^\mu_k}\Psi_{y_k} = 0,\label{BILOC:rel_noR_nonrel}\\
\Box_{y_k}\Psi_{y_k} + \frac{s^2}{y_k^2}\Psi_{y_k} = 0.\label{BILOC:rel_spherical_nonrel}
\end{gather}
For $k^\mu$ timelike the vector $y_k^\mu$ takes values in a three dimensional spacelike hyperplane orthogonal to $k^\mu$. As such \crefrange*{BILOC:rel_noR_nonrel}{BILOC:rel_spherical_nonrel} have the same solution as their non-relativistic counterparts \crefrange*{BILOC:no r}{BILOC:harmonics}, i.e. $\Phi_{y_k}(y_k) = Y^m_\ell$ where $Y^m_\ell$ is a  spherical harmonic. As the Hamiltonian is a sum of the first class constraints this completes the quantization of the relativistic two-particle model. The solutions are characterized by three quantum numbers $M,\ell$ and $m$ where $M \in \R$, $\ell \in \mathbb{N}$ and $m = -\ell, -\ell +1, \ldots, \ell -1 , \ell$; wavefunctions are written as
\begin{align}
\Psi_{M,\ell,m} = \Psi_0 \Psi_k^M Y_\ell ^m,
\end{align}
where $\Psi_0$ is undetermined. 
\section{Conclusion}
In this paper we showed that the relativistic spinning particle can be realized as a bilocal model which itself was explicitly constructed from a constrained non-relativistic system. Such a construction offers insight into the nature of spin, it suggest a deeper relationship between spin and non-locality and deserves further investigation. We were  able to touch on several interesting aspects of the two-particle model namely: the presence of entanglement, the limitation on the total mass for physical constituents and a potential explanation for the nonexistence of higher massive spinning particle above a certain threshold.
We also have seen that constituents carry fractional charges and that our description opens up the possibility of more general coupling to external fields which could exploit the non-locality of the spinning particle. In the body of the paper we considered the case where the constituent particles were of equal mass, only briefly examining the more general case in an Appendix. Some initial investigations described in that Appendix show that in the limit where the total mass vanishes this mass difference is related to the description of continuous spin particles. One of the key open questions for us is to understand whether it is physically possible for spin space to acquire a non-trivial geometry, and whether we can use the framework developed in this paper to generalize curved momentum space models \cite{deepening,scalarQFT} to higher spin fields.
\section*{Acknowledgments}
Research at Perimeter Institute for Theoretical Physics is supported in part by the Government of Canada through NSERC and by the Province of Ontario through MRI. 

\appendix
\section{Fermionic Spherical Harmonics}\label{BILOC:app:harmonics}
In this appendix we include a brief discussion on ``fermionic spherical harmonics'' $Y^m_\ell(\theta, \phi)$ which allow for half-integer values of $m,\ell$, see \cite{fermi_harmonics1,fermi_harmonics2}. We begin with the standard differential equation
\begin{align}\label{BILOC:Gharmonic}
\left[\frac{1}{\sin\theta}\frac{\partial}{\partial \theta}\left(\sin\theta \frac{\partial}{\partial \theta}\right) + \frac{1}{\sin^2\theta}\frac{\partial^2}{\partial\phi^2}\right]Y(\theta, \phi) = -\lam Y(\theta, \phi),
\end{align}
which is separable and we make the assumption that $\lam \geq 0$. Putting $Y(\theta, \phi) = \Theta(\theta)\Phi(\phi)$ we find
\begin{gather}
\sin\theta \frac{d}{d\theta}\left(\sin\theta \frac{d\Theta}{d\theta}\right) + (\lam\sin^2\theta - \kappa)\Theta = 0\label{BILOC:app:theta}\\
\frac{d^2\Phi}{d\phi^2} = -\kappa\Phi
\end{gather}
where $\kappa$ is the separation constant. The second equation is straightforward to solve
\begin{align}
\Phi_m(\phi) = \alpha_1 e^{im\phi} + \alpha_2e^{-im \phi}.
\end{align}
where $m^2 = \kappa$ and $\alpha_1,\alpha_2$ are integration constants. It is standard to argue that $m$ should be an integer since $\phi$ has period $2 \pi$ and $\Phi(\phi)$ must be single valued, however this reasoning is spurious. It is only the  probability density $\abs{\Phi(\phi)}$ which needs to be single valued since it is this quantity which has a physical interpretation. Under this less restrictive assumption we only require that $\Phi_m(\phi)$ is periodic and therefore that $2m \in \mathbb{N}$.\\
\indent Put $\lam = \ell(\ell + 1)$ in equation \eqref{BILOC:app:theta} and make the change of variables $x = \cos\theta$ to obtain
\begin{align}
(1-x^2)\ddot{\Theta} -2x\dot{\Theta} + \left(\ell(\ell + 1) - \frac{m^2}{1-x^2}\right)\Theta = 0,
\end{align}
where a dot indicates a derivative with respect to $x$. Notice that since $\lam$ is assumed to be non-negative $\ell$ is real valued. This is the associated Legendre equation and it's solution is well known, namely $\Theta(x) = \b_1 P^m_\ell(x) + \b_2 Q^m_\ell(x)$ for some constants $\b_1, \b_2$. To have a normalizable wavefunction it is sufficient to require that $\Theta(x)$ be regular on the interval $[-1,1]$; to this end let us examine the behavior of $P^m_\ell(x)$ and $Q^m_\ell(x)$ as $x \to 1^-$. As \cref{BILOC:app:theta} is invariant under $m \to -m$ we can restrict to $m\geq 0$ without loss of generality, we find
\begin{alignat}{2}
P^m_\ell(x) &\sim \left(1-x\right)^{-m/2}, \qquad &m &\neq 1,2,\ldots\label{BILOC:noINTP}\\
P^m_\ell(x) &\sim\left(1-x\right)^{m/2}, \qquad &m &= 1,2,\ldots, \quad \ell - m \neq -1, -2, \ldots \label{BILOC:intP}\\
Q^0_\ell(x) &\sim \log\left(1-x\right), \qquad &\ell & \neq -1,-2,\ldots\label{BILOC:Qzero}\\
Q^m_\ell(x) &\sim \left(1-x\right)^{-m/2}, \qquad &m &\neq \frac{1}{2}, \frac{3}{2}, \ldots\label{BILOC:nohalfQ}\\
Q^m_\ell(x) &\sim \left(1-x\right)^{m/2}, \qquad &m&= \frac{1}{2}, \frac{3}{2},\ldots, \quad \ell - m \neq -1, -2, \ldots\label{BILOC:halfQ}.
\end{alignat}
It follows that a regular solution is only possible if $m$ is either an integer or half-integer, in the former case we have $\Theta(x) = \beta_1 P^m_\ell(x)$ and in the latter $\Theta(x) = \beta_2Q^m_\ell(x)$. The values of $\ell$ are as yet unrestricted, but we still need to consider regularity of the wavefunction as $x \to -1^+$, which can be determined from the following relations
\begin{align}
P^m_\ell(-x) &= \cos((\ell -m)\pi)P^m_\ell(x) - \frac{2}{\pi}\sin((\ell - m)\pi)Q^m_\ell(x).\label{BILOC:minusP}\\
Q^m_\ell(-x) &= -\cos((\ell - m)\pi)Q^m_\ell(x) - \frac{2}{\pi}\sin((\ell -m)\pi)P^m_\ell(x).\label{BILOC:minusQ}.
\end{align}
When $m$ is an integer/half-integer \crefrange{BILOC:noINTP}{BILOC:halfQ} imply that only $P^m_\ell(x)$ respectively $Q^m_\ell(x)$ are finite in the limit $x \to 1^+$. Therefore, if the wavefunction is to be regular as $x \to -1^+$ we require that terms containing the other Legendre function vanish from \cref{BILOC:minusP}/\cref{BILOC:minusQ}. In each case this implies that $\ell - m$ is an integer and so if $m$ is an integer/half-integer $\ell$ is as well. Furthermore, in each case we have that $\ell - m \geq 0$ and since this should be symmetric with respect to $m \to - m$ we also have $\ell +m \geq 0$, combining these conditions gives $-\ell \leq m \leq \ell$. Noting that for $m$ a half-integer $Q^m_\ell(x) \propto P^{-m}_\ell(x)$ we can write the most general solution to \cref{BILOC:app:theta} as
\begin{align}
\Theta^m_\ell(x) &= \beta P^{\eps_\ell \abs{m}}_\ell(x), \qquad \ell = 0,\frac{1}{2},1,\frac{3}{2},\ldots, \quad m = -\ell, -\ell +1, \ldots, \ell -1, \ell\\
\end{align}
where $\eps_\ell = (-1)^{2\ell}$. This result can now be combined with $\Phi_m(\phi)$ to obtain the full solution to \cref{BILOC:Gharmonic} namely $Y^m_\ell(\theta, \phi) = \Theta^m_\ell(\theta)\Phi_m(\phi)$. When $m$ is an integer these are the standard spherical harmonics, however if $m$ is a half-integer we obtain ``fermonic'' spherical harmonics which change sign under $\phi \to \phi +2\pi$. As mentioned earlier, a multivalued wavefunction is acceptable provided that the probability density is single valued and it is easy to verify that this property holds for ``fermonic'' spherical harmonics. 
\section{Unequal Massess}\label{BILOC:app:unequal}
In the non-relativistic model the form of the final Hamiltonian was independent of any mass difference between the constituent particles. This is decidedly not the case when considering the relativistic setting, as will be explored in the current appendix. We begin by defining the masses $M=m_1 +m_2$ and $\mu =m_1m_2/(m_1+m_2)$  and the four-vector coordinates
\begin{align}\label{BILOC:app:rel_var}
\begin{aligned}
X^\mu &= \frac{m_1}{M}x_1^\mu + \frac{m_2}{M} x_2^\mu, \\
P^\mu &= p_1^\mu + p_2^\mu,
\end{aligned}
\qquad\qquad
\begin{aligned}
\Delta x^\mu &= x_1^\mu - x_2^\mu,\\
\Delta p^\mu &= \frac{\mu}{m_1}p_1^\mu - \frac{\mu}{m_2}p_2^\mu,
\end{aligned}
\end{align}
which have Poisson brackets $\pb{X^\mu, P^\nu} = \pb{\Delta x^\mu, \Delta p^\nu} = \eta^{\mu\nu}$ and total angular momenta $J= X\wedge P + \Delta x \wedge \Delta p$. Generalizing the analysis of \Cref{BILOC:sec:rel-twoparticle}, there are two mass-shell constraints
\begin{align}\label{BILOC:mass_shell_uneq}
p_i^2 + m_i^2 = 0, \qquad i = 1,2
\end{align}
both of which must leave $(\Delta x)^2 = \ell^2$ and $(\Delta x \wedge \Delta p) = \hbar^2 s^2$ stationary. Again we find that that the constraints $p_1\cdot \Delta x = p_2\cdot \Delta x = 0$ must be included, and noting that $p_{1}= \frac{m_1}{M} P + \Delta p$ and $p_2 = \frac{m_2}{M} P -\Delta p$ the full Hamiltonian can be written as
\bea
H &= \frac{N}{2 }\left(P^2 +{M}^2+  \frac{M}{\mu}(\Delta p)^2 \right) +
\tilde{N}\left((P\cdot \Delta p) - \frac{\Delta m}{2 \mu}   (\Delta p)^2  \right)
+ \frac{\lam_1}{2}\left((\Delta x)^2 - \ell^2 \right) 
\nonumber  \\  &\qquad 
+ \frac{\lam_2}{2}\left((\Delta p)^2 - \eps^2 s^2\right) + (\lam_3m_1 + \lambda_4 m_2)(P\cdot \Delta x) + (\lambda_3-\lam_4)(\Delta p \cdot \Delta x),
\eea
where we have introduced the mass difference $\Delta m = m_1 -m_2$.


We see that the four constraints 
\be
(P\cdot \Delta x)=0, \quad (\Delta p\cdot \Delta x)=0,\quad (\Delta x)^2=\ell^2, \quad (\Delta p)^2 = \epsilon^2 s^2
\ee 
are identical to the equal mass case, whereas the mass shell and final orthogonality constraint are modified. Specifically, define
\be
\mc{M}^2  := {M}^2 + \frac{M}{\mu}\epsilon^2 s^2,\qquad 
\rho:= \frac{\Delta m }{2\mu}\epsilon^2 s^2,
\ee
then the modified constraints are
\be
P^2 +\mc{M}^2 =0,\qquad (P\cdot \Delta p) = \rho. 
\ee
No further constraints need to be added but demanding that the existing constraints Poisson commute with $H$ imposes the following conditions among the Lagrange multipliers
\begin{gather}
\lam_3 = \lam_4 = 0,  \\
\left(N\frac{M}{\mu} - \tilde{N}\frac{\Delta m}{2 M}+ \lambda_2\right)   
 = \frac{\lambda_1  \ell^2}{\epsilon^2 s^2}= \tilde{N} \frac{\mc{M}^2}{\rho}.
 \end{gather}
It follows that the reduced Hamiltonian involves two unconstrained Lagrange multipliers which correspond to the first class constraints
 \begin{align}
\Phi_P &= P^2 + \mc{M}^2,\\
\Phi_S &=   
\frac{(\Delta p)^2}{\epsilon^2} s^2 
+  \frac{(\Delta x)^2}{\ell^2}  
- 2\hbar^2 s^2  +\frac{\rho}{\mc{M}^2} \left[(P\cdot \Delta p) -  \rho\right].
\end{align}
There are an additional four second class constraints: a modified one $P \cdot \Delta p = \rho$ and three unmodified
\begin{align}
P\cdot \Delta x = 0, \quad \Delta p \cdot \Delta x = 0, \quad \eps^2 s^2 (\Delta x)^2 - \ell^2 (\Delta p)^2 = 0.
\end{align}
The key difference from the equal mass case is the fact that $P\cdot \Delta p \neq 0$ which gives rise to the addtional complexity in the spin cosntraint $\Phi_S$. 

From these expressions it is clear that  the case of continuous spin particles\footnote{The idea of continuous spin particles in the DPS framework will be discussed more fully in future work.} \cite{wigner_1937,edgren_2005,schuster_2013A} can then be obtained in the limit where $\mc{M} \to 0$ while keeping $\rho $ fixed. Indeed, in this limit we recover the constraints 
\be
P^2=0,\quad  P\cdot \Delta x = 0,  \quad P \cdot \Delta p = \rho
\ee
together with
$\eps^2 s^2 (\Delta x)^2 + \ell^2 (\Delta p)^2= 2\hbar^2 s^2$ and $\Delta p \cdot \Delta x = 0$, $\eps^2 s^2 (\Delta x)^2 = \ell^2 (\Delta p)^2$. These are the constraints for a continuous spin particle.
\\
\indent At the outset of this appendix we put $X_\mu$ as the ``center of mass'' but this choice was arbitrary. 
Another option is to look for a definition of $X'$ which leads to a vanishing mixing parameter $\rho$. Note that in order to keep the canonical algebra, changing $X$ also means that we are changing $\Delta p$.
Lets consider 
\be 
X' = X - \frac{\Delta m}{2 \mu} \frac{\epsilon^2 s^2}{\mc{M}^2} P, \qquad \Delta p' = \Delta p + \frac{\Delta m}{2 \mu} \frac{\epsilon^2 s^2}{\mc{M}^2} P,
\ee
which preserve the canonical algebra by construction and satisfy
$P\cdot \Delta p' = 0$. 
This change of coordinates can be seen as a redefinition of the effective spin, which is now given by $\epsilon^2 s'^2 = (\Delta p')^2$, while also rendering the position coordinate $X'$ momentum dependent. For example, imagine coupling the massive spinning particle to an external electromagnetic field: With a vanishing mixing parameter it is natural to consider the coupling $A(X')$, however when expressed in the CSP frame where the mixing doesn't vanish this reads $A(X+ \a P)$ and the location of the coupling is now momentum dependent.

\bibliographystyle{ieeetr}
\bibliography{thesis}{}

\begin{thebibliography}{10}

\bibitem{yukawa_1949A}
H.~Yukawa, ``{Quantum Theory of Nonlocal Fields. 1. Free Fields},'' {\em
  Phys.Rev.}, vol.~77, pp.~219--226, 1950.

\bibitem{yukawa_1950B}
H.~Yukawa, ``{Quantum Theory of Nonlocal Fields. 2: Irreducible Fields and
  Their Interaction},'' {\em Phys.Rev.}, vol.~80, pp.~1047--1052, 1950.

\bibitem{takabayasi_1979}
T.~Takabayasi, ``{Relativistic Mechanics of Confined Particles as Extended
  Model of Hadrons: The Bilocal Case},'' {\em Prog.Theor.Phys.Suppl.}, vol.~67,
  p.~1, 1979.

\bibitem{two_interacting_relativistic}
D.~Dominici, J.~Gomis, and G.~Longhi, ``A lagrangian for two interacting
  relativistic particles,'' {\em Il Nuovo Cimento B (1971-1996)}, vol.~48,
  no.~2, pp.~152--166, 1978.

\bibitem{second_quantization_nonhadrons}
R.~Casalbuoni, D.~Dominici, and G.~Longhi, ``On the second quantization of a
  composite model for nonhadrons,'' {\em Il Nuovo Cimento A (1965-1970)},
  vol.~32, no.~3, pp.~265--275, 1976.

\bibitem{bilocal_and_string}
T.~Goto, S.~Naka, and K.~Kamimura, ``On the bi-local model and string model,''
  {\em Supplement of the Progress of theoretical physics}, pp.~69--114, apr
  1980.

\bibitem{mechanical_higher_spin}
A.~K.~H. Bengtsson, ``{Mechanical Models for Higher Spin Gauge Fields},'' {\em
  Fortsch. Phys.}, vol.~57, pp.~499--504, 2009.

\bibitem{rempel_2015}
T.~Rempel and L.~Freidel, ``{Interaction Vertex for Classical Spinning
  Particles},'' 2015.

\bibitem{kirillov_2002}
A.~Kirillov, {\em {Lectures on the Orbit Method}}.
\newblock American Mathematical Society., 2004.

\bibitem{non_rel_model}
A.~A. {Deriglazov} and A.~M. {Pupasov-Maksimov}, ``{Geometric Constructions
  Underlying Relativistic Description of Spin on the Base of Non-Grassmann
  Vector-Like Variable},'' {\em SIGMA}, vol.~10, p.~12, Feb. 2014.

\bibitem{fermi_harmonics1}
G.~{Hunter}, P.~{Ecimovic}, I.~{Schlifer}, I.~M. {Walker}, D.~{Beamish},
  S.~{Donev}, M.~{Kowalski}, S.~{Arslan}, and S.~{Heck}, ``{Fermion
  quasi-spherical harmonics},'' {\em Journal of Physics A Mathematical
  General}, vol.~32, pp.~795--803, Feb. 1999.

\bibitem{fermi_harmonics2}
G.~{Hunter} and M.~{Emami-Razavi}, ``{Properties of Fermion Spherical
  Harmonics},'' {\em eprint arXiv:quant-ph/0507006}, July 2005.

\bibitem{single_valued}
E.~{Merzbacher}, ``{Single Valuedness of Wave Functions},'' {\em American
  Journal of Physics}, vol.~30, pp.~237--247, Apr. 1962.

\bibitem{costa_2012}
L.~Costa, C.~Herdeiro, J.~Nat\'ario, and M.~Zilh\~ao, ``Mathisson's helical
  motions for a spinning particle: Are they unphysical?,'' {\em Phys. Rev. D},
  vol.~85, p.~024001, Jan 2012.

\bibitem{costa_2014}
L.~F. Costa and J.~Nat{\'a}rio, ``{Center of mass, spin supplementary
  conditions, and the momentum of spinning particles},'' 2014.

\bibitem{deepening}
G.~Amelino-Camelia, L.~Freidel, J.~Kowalski-Glikman, and L.~Smolin, ``{Relative
  locality: A deepening of the relativity principle},'' {\em Gen.Rel.Grav.},
  vol.~43, pp.~2547--2553, 2011.

\bibitem{scalarQFT}
L.~Freidel and T.~Rempel, ``{Scalar Field Theory in Curved Momentum Space},''
  2013.

\bibitem{wigner_1937}
E.~P. Wigner, ``{On Unitary Representations of the Inhomogeneous Lorentz
  Group},'' {\em Annals Math.}, vol.~40, pp.~149--204, 1939.

\bibitem{edgren_2005}
L.~Edgren, R.~Marnelius, and P.~Salomonson, ``{Infinite spin particles},'' {\em
  JHEP}, vol.~0505, p.~002, 2005.

\bibitem{schuster_2013A}
P.~Schuster and N.~Toro, ``{On the Theory of Continuous-Spin Particles:
  Wavefunctions and Soft-Factor Scattering Amplitudes},'' {\em JHEP},
  vol.~1309, p.~104, 2013.

\end{thebibliography}
\end{document}